%% file: Incera-CSQCD.tex
\documentclass[12pt]{article}
\usepackage{epsfig}

\input econfmacros.tex
\def\Title#1{\begin{center} {\Large {\bf #1} } \end{center}}

\begin{document}

\Title{Cold and Dense Matter in a Magnetic Field}

\bigskip\bigskip


\begin{raggedright}

{\it Vivian de la Incera\index{de la Incera, V.}\\
Department of Physics\\
University of Texas at El Paso\\
500 W. University Ave.\\
El Paso, TX 79968\\
USA\\
{\tt Email: vincera@utep.edu}}
\bigskip\bigskip
\end{raggedright}

\begin{abstract}

Our Universe is full of regions where extreme physical
conditions are realized. Among the most intriguing cases
are the so-called magnetars: neutron stars with very dense
cores and super-strong magnetic fields. In this paper I 
review the current understanding of the physical properties
of quark matter at ultra-high density in the presence of
very large magnetic fields. I will discuss the main results
on this topic, the main challenges that still remain, and
how they could be related to the physics of magnetars.
\end{abstract}

\section{Introduction}

The realm of high density QCD pertains to situations of very high baryon density. At
very high baryon density (and low temperatures) baryons get so squeezed that
they start to overlap, thereby erasing any vestige of structure. Since in
this case the quarks get very close to each other, the phenomenon of asymptotic freedom
ensures their interactions to become weaker and weaker when the density becomes
higher and higher. At densities of the order
of 10 times the nuclear density the quarks will be
so weakly interacting that they can exist out of
confinement. What is particularly interesting about cold and dense quark matter is that
the fundamental QCD interaction is attractive in the color
antitriplet channel. Once the quarks are deconfined and fill out the available quantum states up to the Fermi surface, this attractive interaction triggers the formation of diquark pairs at the Fermi surface, thus leading to the phenomenon of color superconductivity (CS) (for a historical account and detailed discussion of CS see \cite{reviews}).

In nature the combination of the high densities and relatively
low temperatures required for color superconductivity can be found in the interior
of neutron stars, which are the remnant
of supernova explosions. At the same time, it is well-known
\cite{Grasso} that strong magnetic fields, as large as $B \sim
10^{12} - 10^{13}$ G,  exist in the surface of regular neutron stars, while
in the case of magnetars they are in the range
$B \sim 10^{14} - 10^{15}$ G, and
perhaps as high as $10^{16}$ G \cite{magnetars}. Moreover, the virial
theorem \cite{virial} allows the field magnitude to reach values as
large as $10^{18}-10^{19}$ G. To produce reliably predictions of astrophysical
signatures of color superconductivity, a better understanding of the
role of the star's magnetic field in the color superconducting (CS) phase
is essential.

In recent years, several works \cite{MCFL1}-\cite{MPhases} have been
dedicated to elucidate the influence of a magnetic field in the ground
state of CS matter. These investigations have revealed a richness of
phases \cite{MPhases} with different symmetries and low energy
properties.

In order to grasp how a magnetic field can affect the color
superconducting pairing, it is important to recall that in spin-zero
color superconductivity, although the color condensate has non-zero
electric charge, a linear combination of the photon and one of the
gluons remains massless \cite{alf-raj-wil-99/537,
alf-berg-raj-NPB-02}, so the condensate is neutral with respect to the
Abelian charge associated with the symmetry group of this long-range
gauge field. This combination then behaves as the
"in-medium" (also called "rotated") electromagnetic field in the color superconductor. Since
this combination acquires no mass, there is no Meissner effect for the
corresponding "rotated" magnetic field and consequently, a spin-zero color superconductor
may be penetrated by a rotated magnetic field
$\widetilde{B}$. Moreover, it is worth to notice that despite
all the superconducting pairs are neutral
with respect to this long-range field, a subset of them is formed by
quarks of opposite rotated charges ${\widetilde Q}$. The interaction of the charged
quarks with the magnetic field  gives rise to a difference between the gaps getting contribution from pairs
formed by oppositely charged quarks and those getting contribution only from pairs of neutral quarks. One consequence of such a difference is the change of the gap parameter symmetry \cite{MCFL1}. If the field is strong enough, it actually strengthens the pairing of quarks of oppositely rotated charge \cite{MCFL2}. One can intuitively understand this considering that the quarks with opposite charges ${\widetilde Q}$ and opposite spins, have parallel (rather
than antiparallel) magnetic moments, so the field tends to keep the alignment of
these magnetic moments, hence helping to stabilize the pairing of these quarks.

Besides changing the symmetry of the gap and consequently the low-energy
physics, a magnetic field can lead to other interesting behaviors too. It can produce oscillations in the gaps and the magnetization \cite{Moscillations}, the Hass-Van
Alphen effect. Moreover, when the field strength is of the order of the Meissner mass of the rotated charged gluons, these modes become tachyonic \cite{PCFL1}-\cite{PCFL2}. The solution to this instability is the formation of a vortex state of gluons which in turn boosts the magnetic field, creating a peculiar paramagnetic state. This magnetic-field induced gluon vortex state is known as the Paramagnetic CFL (PCFL) phase.

The paper is organized as follows. Section 2 and 3 outline the effects of a magnetic field in the gap magnitude and structure for three- and two-flavor spin zero color superconductors. In Section 4 I briefly mention several questions derived from the results here presented, as well as the main standing problems in the field of color superconductivity and how taking into account the magnetic field (whether external or induced by the color superconductor) we could address some of them and have as a byproduct a potential solution for some of the puzzles in the physics of magnetars.

\section{Three Quark Flavors in a Magnetic Field}
\subsection{MCFL Symmetry}

Let us consider a model of three quark flavors in the presence of a magnetic field at high baryon density. This system was investigated in \cite{MCFL1} in the context of a NJL model based on the one-gluon exchange interaction of QCD. The first important thing to notice in this case is that a magnetic field affects the flavor symmetries of QCD, as different quark flavors have different electromagnetic charges. For
three light quark flavors, only the subgroup of $SU(3)_L \times
SU(3)_R$ that commutes with $Q$, the electromagnetic charge operator, is a
symmetry of the theory. Based on the above considerations, and imposing that in the presence of an external magnetic field the
condensate should retain the highest degree of symmetry, one can propose
\cite{MCFL1} the following ansatz for the gap structure in the
presence of a magnetic field

\begin{equation}
\Delta=\left(
\begin{array}{ccccccccc}
0 & 0 & 0 & 0 & \Delta_{A} & 0 & 0 & 0 & \Delta_{A}^{B}\\
0 & 0 & 0 & -\Delta_{A} & 0 & 0 & 0 & 0& 0 \\
0 & 0 & 0 & 0 & 0 & 0 & -\Delta_{A}^{B}& 0 & 0 \\
0 & -\Delta_{A}& 0 & 0 & 0 & 0 & 0 & 0 & 0 \\
\Delta_{A}& 0 & 0 & 0 & 0 & 0 & 0 & 0 &  \Delta_{A}^{B}\\
0 & 0 & 0 & 0 & 0 & 0& 0 & -\Delta_{A}^{B} & 0 \\
0 & 0 & -\Delta_{A}^{B}& 0 & 0 & 0 & 0 & 0 & 0 \\
0 & 0 & 0 & 0 & 0 & -
\Delta_{A}^{B}& 0 & 0&0\\
\Delta_{A}^{B} & 0 & 0 & 0 &
\Delta_{A}^{B} & 0 & 0 &0&0
\label{order-parameter}
\end{array} \right)
\end{equation}
 
The above gap is based on the quark representation
$\psi^{T}=(s_{1},s_{2},s_{3},d_{1},d_{2},d_{3},u_{1},u_{2},u_{3})$.
Notice that we ignored the symmetric gaps (see \cite{MCFL2} for the complete set of symmetric and antisymmetric gaps in the gap structure in a magnetic field). For the purpose of symmetry considerations, they are no relevant, as they do not break any symmetry that is not already broken by the antisymmetric gaps. However, the reader should be aware that the symmetric gaps are nonzero. In general they are smaller than the antisymmetric gaps, because they originate from a color-repulsive, rather than attractive interaction. Nevertheless, we call attention that at magnetic fields of the order or larger than the baryon chemical potential scale, the symmetric gap that gets contributions from pairs of charged quarks can be of the order or larger than the antisymmetric gap that only gets contributions from pairs of neutral quarks
\cite{MCFL2}.

The order parameter (\ref{order-parameter}) implies the
following symmetry breaking pattern: 
\begin{equation}
SU(3)_{\rm color} \times
SU(2)_L \times SU(2)_R \times U(1)_B \times U^{(-)}(1)_A \rightarrow SU(2)_{{\rm color}+L+R}.
\end{equation}
The $U^{(-)}(1)_A$ symmetry is
connected with the current which is an anomaly-free linear
combination of $s,d$ and $u$ axial currents
\cite{miransky-shovkovy-02}.  The locked $SU(2)$ corresponds to the
maximal unbroken symmetry, and as such it maximizes the condensation
energy. Given that it commutes with the rotated electromagnetic
group ${\widetilde {U}(1)}_{\rm e.m.}$, the rotated electromagnetism remains as a symmetry of the MCFL phase.

The phase described by the order parameter (\ref{order-parameter}) is known in the literature as the Magnetic CFL (MCFL) phase. It locks SU(2) left and right flavor transformations with SU(3) color transformations, similar to the CFL phase, so it also breaks the chiral symmetry of the original theory. The low-energy physics of the MCFL phase differs from the CFL one, as it is characterized by five instead of nine
Nambu-Goldstone (NG) bosons. One of the NG bosons is associated to the breaking of the
baryon symmetry; three others are associated to the
breaking of $SU(2)_A$, and another one is associated to the breaking of
$U^{(-)}(1)_A$. The propagation of light in the MCFL phase is also different from the CFL case, as all of
the five NG bosons in the MCFL phase are
$\widetilde{Q}$-neutral.

\subsection{MCFL Effective Action}

The effective action of the three-flavor quark system in the presence
of a magnetic field can be obtained using a Nambu-Jona-Lasinio
(NJL) Lagrangian with the four-fermion interaction abstracted from one-gluon exchange
\cite{alf-raj-wil-99/537}.

The mean-field effective action for such a theory can be written as
\begin{eqnarray}
I_{B}(\overline{\psi},\psi )
=\int\limits_{x,y}\{\frac{1}{2}[\overline{\psi }
_{(0)}(x)[G_{(0)0}^{+}]^{-1}(x,y)\psi _{(0)}(y)+\overline{ \psi
}_{(+)}(x)[G_{(+)0}^{+}]^{-1}(x,y)\psi _{(+)}(y) \nonumber \\
+\overline{\psi}_{(-)}(x)[G_{(-)0}^{+}]^{-1}(x,y)\psi _{(-)}(y)+
\overline{\psi } _{(0)C}(x)[G_{(0)0}^{-}]^{-1}(x,y)\psi
_{(0)C}(y)\nonumber \\
+\overline{ \psi }_{(+)C}(x)[G_{(+)0}^{-}]^{-1}(x,y)\psi _{(+)C}(y)
+\overline{\psi}_{(-)C}(x)[G_{(-)0}^{-}]^{-1}(x,y)\psi _{(-)C}(y)]
\nonumber \\
+\frac{1}{2}[\overline{\psi }_{(0)C}(x)\Delta ^{+}(x,y)\psi
_{(0)}(y)+h.c.] + \frac{1}{2}[\overline{\psi }_{(+)C}(x)\Delta
^{+}(x,y)\psi _{(-)}(y)\nonumber
\\
+\overline{\psi }_{(-)C}(x)\Delta ^{+}(x,y)\psi _{(+)}(y)+h.c.]\} \
, \label{b-coord-action}
\end{eqnarray}
where the external magnetic field has been explicitly introduced
through minimal coupling with the $\widetilde{Q}-$charged fermions.
The presence of the field is also taken into account in the diquark
condensate $\Delta^{+}=\gamma_{5}\Delta$, whose color-flavor
structure is given by Eq.(\ref{order-parameter}).

In (\ref{b-coord-action}) symbols in parentheses indicate neutral
$(0)$, positive $(+)$ or negative $(-)$ $\tilde{Q}-$charged quarks.
Supra-indexes $+$ or $-$ in the propagators indicate, as it is
customary, whether it is the inverse propagator of a field or
conjugated field respectively. Then, for example,
$[G_{(+)0}^{-}]^{-1}$ corresponds to the bare inverse propagator of
positively charged conjugate fields, and so on. The explicit
expressions of the inverse propagators are

\begin{equation}
\lbrack G_{(0)0}^{\pm}]^{-1}(x,y)=[i\gamma ^{\mu }\partial _{\mu
}-m\pm \mu \gamma ^{0}]\delta ^{4}(x-y) \ ,  \label{neut-x-inv-prop}
\end{equation}

\begin{equation}
\lbrack G_{(+)0}^{\pm }]^{-1}(x,y)=[i\gamma ^{\mu }\Pi ^{(+)}_{\mu
}-m\pm \mu \gamma ^{0}]\delta ^{4}(x-y) \ ,  \label{B-x-inv-prop+}
\end{equation}

\begin{equation}
\lbrack G_{(-)0}^{\pm }]^{-1}(x,y)=[i\gamma ^{\mu }\Pi ^{(-)}_{\mu
}-m\pm \mu \gamma ^{0}]\delta ^{4}(x-y) \ ,  \label{B-x-inv-prop-}
\end{equation}
with

\begin{equation}
\Pi ^{(\pm)}_{\mu }=i\partial _{\mu
}\pm\widetilde{e}\widetilde{A}_{\mu } \ . \label{piplusminus}
\end{equation}

Transforming the field-dependent quark propagators to momentum space
can be performed with the use of the Ritus' method, originally
developed for charged fermions \cite{Ritus:1978cj} and later
extended to charged vector fields \cite{efi-ext}. In Ritus' approach
the diagonalization in momentum space of charged fermion Green's
functions in the presence of a background magnetic field is carried
out using the eigenfunction matrices $E_p(x)$. These are the wave
functions of the asymptotic states of charged fermions in a uniform
magnetic field and play the role in the magnetized medium of the
usual plane-wave (Fourier) functions $e^{i px}$ at zero field.

The transformation functions $E^{(\pm)}_{q}(x)$ for positively
($+$), and negatively ($-$) charged fermion fields are obtained as
the solutions of the field dependent eigenvalue equation

\begin{equation}
(\Pi^{(\pm)}\cdot\gamma)E^{(\pm)}_{q}(x)=E^{(\pm)}_{q}(x)(\gamma\cdot\overline{p}^{(\pm)}),
\label{eigenproblem}
\end{equation}
with $\overline{p}^{(\pm )}$ given by

\begin{equation}  \label{pbar+}
\overline{p}^{(\pm )}=(p_{0},0,\pm
\sqrt{2|\widetilde{e}\widetilde{B}|k},p_{3}) \ ,
\end{equation}
and
\begin{equation}
E^{(\pm)}_{q}(x)=\sum\limits_{\sigma }E^{(\pm)}_{q\sigma }(x)\delta
(\sigma ) \ , \label{9}
\end{equation}
with eigenfunctions

\begin{equation}
{E}^{(\pm)}_{p\sigma }(x)=\mathcal{N}
_{n_{(\pm)}}e^{-i(p_{0}x^{0}+p_{2}x^{2}+p_{3}x^{3})}D_{n_{(\pm)}}(\varrho
_{(\pm)})  ,
 \label{Epsigma+}
\end{equation}
where $D_{n_{(\pm)}}(\varrho _{(\pm)})$ are the parabolic cylinder
functions with argument $\varrho _{(\pm)}$ defined by
\begin{equation}
\varrho _{(\pm)}=\sqrt{2|\widetilde{e}\widetilde{B}|}(x_{1}\pm p_{2}/\widetilde{e}
\widetilde{B}) \ ,
 \label{rho+}
\end{equation}
and index $n_{(\pm)}$ given by
\begin{equation}  \label{normaliz-const+}
n_{(\pm)}\equiv n_{(\pm)}(k,\sigma)= k \pm \frac{\widetilde{e}\widetilde{B}}{2|
\widetilde{e}\widetilde{B}|}\sigma-\frac{1}{2} \ , \qquad\qquad
n_{(\pm)}=0,1,2,...
\end{equation}
$k=0,1,2,3,...$ is the Landau level, and $\sigma$ is the spin
projection that can take values $\pm 1$ only. Notice that in the
lowest Landau level, $k=0$, only particles with one of the two spin
projections, namely, $\sigma =1$ for
positively charged particles, are allowed. The normalization constant $
\mathcal{N}_{n_{(\pm)}}$ is
\begin{equation}  \label{normaliz-const}
\mathcal{N}_{n_{(\pm)}}=(4\pi| \widetilde{e}\widetilde{B}|)^{\frac{1}{4}}/
\sqrt{n_{(\pm)}!} \ .
\end{equation}

The spin matrices $\delta(\sigma )$ are defined as

\begin{equation}
\delta(\sigma )= {\rm diag}(\delta _{\sigma 1},\delta _{\sigma
-1},\delta _{\sigma 1},\delta _{\sigma -1}),\qquad \sigma =\pm 1 \ ,
\label{10}
\end{equation}
and satisfy the following relations

\begin{equation}
\delta \left( \pm \right) ^{\dagger }=\delta \left( \pm \right)
,\qquad \delta \left( \pm \right) \delta \left( \pm \right) =\delta
\left( \pm \right)  \ ,\qquad \delta \left( \pm \right) \delta
\left( \mp \right) =0
 ,
\end{equation}

\begin{equation}
\gamma ^{\|}\delta \left( \pm \right) =\delta \left(
\pm \right) \gamma ^{\| },\quad \gamma ^{\bot }\delta
\left( \pm \right) =\delta \left( \mp \right) \gamma ^{\bot } \ .
\label{24}
\end{equation}
In Eq. (\ref{24}) the notation $\gamma ^{\|}=(\gamma
^{0},\gamma ^{3})$ and $\gamma ^{\bot }=(\gamma ^{1},\gamma ^{2})$
was used.

 The functions $E^{(\pm)}_{p}$ are complete
\begin{equation}  \label{complete-Ep}
\sum_{k}\int dp_{0}dp_{2}dp_{3}{E}^{(\pm)}_{p}(x){\overline{E}}
^{(\pm)}_{p}(y)=(2\pi)^{4}\delta^{(4)}(x-y) \ ,
\end{equation}
and orthonormal,
\begin{equation}  \label{ortho-Ep}
\int_{x}{\overline{E}}^{(\pm)}_{p^{\prime}}(x){E}^{(\pm)}_{p}(x)=(2\pi)^{4}
\Lambda_{k}\delta_{kk^{\prime}}\delta(p_{0}-p^{\prime}_{0})
\delta(p_{2}-p^{\prime}_{2})\delta(p_{3}-p^{\prime}_{3})
\end{equation}
with the $(4\times4)$ matrix $\Lambda_{k}$ given by

\begin{eqnarray}
\Lambda_{k}= \left\{
\begin{array}{cc}
\delta(\sigma= {\rm sgn}[eB]) \qquad\qquad for \qquad  k=0,\\
\qquad I \qquad\qquad\qquad\qquad for \qquad k>0 .
\end{array} \right.
\end{eqnarray}

In Eqs. (\ref{complete-Ep})-(\ref{ortho-Ep}) we
introduced the notation $\overline{E}_{p}^{(\pm)}(x)=\gamma
_{0}({E}_{p}^{(\pm)}(x))^{\dag }\gamma _{0}$.

Under the $E_p(x)$ functions, positively ($\psi _{(+)}$), negatively
($\psi _{(-)}$) charged fields transform according to
\begin{equation}
\psi _{(\pm)}(x)=\sum_{k}\int dp_{0}dp_{2}dp_{3}E_{p}^{(\pm)}(x)\psi
_{(\pm)}(p) \ ,\label{psi+-transf}
\end{equation}
\begin{equation}
\overline{\psi }_{(\pm)}(x)=\sum_{k}\int dp_{0}dp_{2}dp_{3}\overline{\psi }
_{(\pm)}(p)\overline{E}_{p}^{(\pm)}(x)  \ . \label{psibar+-transf}
\end{equation}

One can show that
\begin{equation}  \label{gamma-piplus-propert}
[\gamma_{\mu}(\Pi_{(+)\mu}\pm \mu\delta_{\mu0})- m]{E}^{(+)}_{p}(x)={E}
^{(+)}_{p}(x)[\gamma_{\mu}(\overline{p}^{(+)}_{\mu}\pm
\mu\delta_{\mu0})- m] \ ,
\end{equation}
and
\begin{equation}  \label{gamma-piminus-propert}
[\gamma_{\mu}(\Pi_{(-)\mu}\pm \mu\delta_{\mu0})- m]{E}^{(-)}_{p}(x)={E}
^{(-)}_{p}(x)[\gamma_{\mu}(\overline{p}^{(-)}_{\mu}\pm
\mu\delta_{\mu0})- m] \ .
\end{equation}

The conjugate fields transform according to,

\begin{equation}  \label{conjpsiplustransf}
\psi_{(+)C}(x)=\sum_{k}\int
dp_{0}dp_{2}dp_{3}E^{(-)}_{p}(x)\psi_{(+)C}(p),
\end{equation}
\begin{equation}  \label{conjpsiminustransf}
\psi_{(-)C}(x)=\sum_{k}\int
dp_{0}dp_{2}dp_{3}E^{(+)}_{p}(x)\psi_{(-)C}(p) \ .
\end{equation}

After transforming to momentum space one can introduce Nambu-Gorkov
fermion fields of different $\tilde{Q}$ charges. They are the
$\tilde{Q}$-neutral Gorkov field

\begin{equation}
\Psi _{(0)}=\left(
\begin{array}{c}
\psi _{(0)} \\
\psi _{(0)C}
\end{array}
\right) \ ,
\end{equation}
the positive
\begin{equation}
\Psi _{(+)}=\left(
\begin{array}{c}
\psi _{(+)} \\
\psi _{(-)C}
\end{array}
\right) \ ,
\end{equation}
and the negative one
\begin{equation}
\Psi _{(-)}=\left(
\begin{array}{c}
\psi _{(-)} \\
\psi _{(+)C}
\end{array}
\right) \ .
\end{equation}

Using them, the Nambu-Gorkov effective action in the presence of a
constant magnetic field $\widetilde{B}$ can be written as

\begin{eqnarray}  \label{b-action}
I^{B}(\overline{\psi},\psi)
=\frac{1}{2}\int\frac{d^{4}p}{(2\pi)^{4}} \overline{\Psi}%
_{(0)}(p){\cal S}^{-1}_{(0)}(p)\Psi_{(0)}(p)\nonumber\\
+\frac{1}{2}\int\frac{d^{4}p}{(2\pi)^{4}}
\overline{\Psi}_{(+)}(p){\cal S}^{-1}_{(+)}(p)\Psi_{(+)}(p)+\frac{1}{2}\int\frac{d^{4}p}{%
(2\pi)^{4}} \overline{\Psi}_{(-)}(p){\cal
S}^{-1}_{(-)}(p)\Psi_{(-)}(p) \ ,
\end{eqnarray}
where
\begin{eqnarray}  \label{p-neutr-inv-propg}
{\cal S}^{-1}_{(0)}(p)=\left(
\begin{array}{cc}
[G_{(0)0}^{+}]^{-1}(p) & \Delta_{(0)}^{-} \\
&  \\
\Delta_{(0)}^{+} & [G_{(0)0}^{-}]^{-1}(p)
\end{array}
\right) \ ,
\end{eqnarray}

\begin{eqnarray}  \label{p-posit-inv-propg}
{\cal S}^{-1}_{(+)}(p)=\left(
\begin{array}{cc}
[G_{(+)0}^{+}]^{-1}(p) & \Delta_{(+)}^{-} \\
&  \\
\Delta_{(+)}^{+} & [G_{(+)0}^{-}]^{-1}(p)
\end{array}
\right) \ ,
\end{eqnarray}
\begin{eqnarray}  \label{p-negat-inv-propg}
{\cal S}^{-1}_{(-)}(p)=\left(
\begin{array}{cc}
[G_{(-)0}^{+}]^{-1}(p) & \Delta_{(-)}^{-} \\
&  \\
\Delta_{(-)}^{+} & [G_{(-)0}^{-}]^{-1}(p)
\end{array}
\right) \ ,
\end{eqnarray}
with

\begin{equation}
\Delta_{(+)}^{+}= \Omega_{-}\Delta^{+}\Omega_{+},
\end{equation}

\begin{equation}
\Delta_{(-)}^{+}= \Omega_{+}\Delta^{+}\Omega_{-},
\end{equation}

\begin{equation}
\Delta_{(0)}^{+}= \Omega_{0}\Delta^{+}\Omega_{0},
\end{equation}

Here we introduced the rotated charge-projector operators

\begin{equation}
\Omega _{0}={\rm diag}(1,1,0,1,1,0,0,0,1) \ ,
 \label{neut-omeg}
\end{equation}

\begin{equation}
\Omega _{+}={\rm diag}(0,0,0,0,0,0,1,1,0) \ ,
\label{pos-omeg}
\end{equation}

\begin{equation}
\Omega _{-}={\rm diag}(0,0,1,0,0,1,0,0,0)  \ ,
\label{neg-omeg}
\end{equation}
which obey the algebra

\begin{equation}
\Omega _{\eta }\Omega _{\eta ^{\prime }}=\delta _{\eta \eta ^{\prime
}}\Omega _{\eta },\qquad \eta ,\eta ^{\prime }=0,+,- \ .
\label{omeg-algeb}
\end{equation}

\begin{equation}
\Omega _{0}+\Omega _{+}+\Omega _{-}=1 \ .
\label{omeg-sum}
\end{equation}

With the help of the charge projectors
one can write the rotated charge operator and the $(0)$-, ($+/-)$-charged fields as
\begin{equation}
\widetilde{Q}=\sum\limits_{\eta =0,\pm }\eta \Omega _{\eta }=\Omega
_{+}-\Omega _{-}  \ .
 \label{Q-omeg-relat}
\end{equation}
and
\begin{equation}
\psi _{0}=\Omega _{0}\psi  \ ,\qquad \psi _{+}=\Omega _{+}\psi  \ ,\qquad
\psi _{-}=\Omega _{-}\psi \ .  \label{fields-def}
\end{equation}
respectively.

Notice that to form the positive (negative) Nambu-Gorkov field we
used the positive (negative) fermion field and the charge conjugate
of the negative (positive) field. This is done so that the rotated
charge of the up and down components in a given Nambu-Gorkov field
be the same. This way to form the Nambu-Gorkov fields is mandated by
what kind of field enters in a given condensate term, which in turn
is related to the neutrality of the fermion condensate $\langle
\overline{\psi}_{C}\psi\rangle$ with respect to the rotated
$\tilde{Q}$-charge.

In momentum space the bare inverse propagator for the neutral field
is

\begin{equation}
\lbrack G_{(0)0}^{\pm }]^{-1}(p)=[\gamma _{\mu }(p_{\mu }\pm \mu
\delta _{\mu 0})-m] \ , \label{neut-bareprop+-}
\end{equation}
where the momentum is the usual $p=(p_{0},p_{1},p_{2},p_{3})$ of the
case with no background field.

For positively and negatively charged fields the bare inverse
propagators are
\begin{equation}
\lbrack G_{(+)0}^{\pm }]^{-1}(p)=[\gamma _{\mu }(\overline{p}
_{\mu }^{(+)}\pm \mu \delta _{\mu 0})-m] \ , \label{pos-bareprop}
\end{equation}
and
\begin{equation}
\lbrack G_{(-)0}^{\pm }]^{-1}(p)=[\gamma _{\mu }(\overline{p}
_{\mu }^{(-)}\pm \mu \delta _{\mu 0})-m]  \ \label{neg-bareprop}
\end{equation}
respectively.

\subsection{MCFL at Strong Magnetic Fields}

The details of the gap equations obtained from (\ref{b-coord-action}) can be found in \cite{MCFL2}. In the strong field limit $\widetilde{e}\widetilde{B} \sim \mu^2$, they become

\begin{equation} \label{maingeq} \Delta^B_A  \approx  \frac{g^2}{3 \Lambda^2}
\int_{\Lambda} \frac{d^3 q}{(2 \pi)^3} \frac{
\Delta^B_A}{\sqrt{(q-\mu)^2 + 2 (\Delta^B_A)^2 }}
 +  \frac{g^2 \widetilde{e}\widetilde{B}}{3 \Lambda^2} \int_{-\Lambda}^{\Lambda}
\frac{d q}{(2 \pi)^2} \frac{ \Delta^B_A}{\sqrt{(q-\mu)^2 +
(\Delta^B_A)^2 }}   \ ,
\end{equation}

\begin{equation} \label{antisymme-gapeqapp}
 \Delta_A  \approx  \frac {g^2}{4 \Lambda^2}  \int_{\Lambda} \frac{d^3q}{(2 \pi)^3}
 \Big( \frac {17}{9} \frac{\Delta_A}{ \sqrt{ (q - \mu)^2 - \Delta_A^2 }}
 +   \frac{7}{9} \frac{\Delta_A}{ \sqrt{(q-\mu)^2 + 2 (\Delta^B_A)^2 }  }
 \Big) \ ,
 \end{equation}

with solution
\begin{equation}\label{mCFLgap}
\Delta_{A}^{B}\sim2\sqrt{\Lambda\mu-\mu^{2}} \mbox{ exp}\left(-\frac{3\Lambda^{2}\pi^{2}}{g^{2}\left(\mu^{2}+\frac{\tilde{e}\tilde{B}}{2}\right)}\right).
\end{equation}
for the gap receiving contribution from pairs of charged quarks. It is instructive to look at the form of this gap. Just as in the conventional BCS gap, this strong-field MCFL gap goes as exp$\left(-1/N \tilde{G}\right)$, where $\tilde{G}=g^{2}/3\Lambda^{2}$ is the dimensionful coupling constant, and $N$ represents the total density of states at the Fermi surface of those quarks contributing to the gap. In the absence of a magnetic field, the density of states for a single quark is $N_{\mu}=\mu^{2}/2\pi^{2}$. The total density of states is then $N=4N_{\mu}$, as there are four quarks lending to each gap. When a magnetic field is present, the contributing quarks are shared among two terms, $N_{\mu}$ and $N_{\tilde{B}}$, depending on whether they are neutral or charged. Here, $N_{\tilde{B}}=\tilde{e}\tilde{B}/4\pi^{2}$. In the present case, we have $N=2N_{\mu}+2N_{\tilde{B}}$, as two of the four quarks are charged. We then have the above expression (\ref{mCFLgap}).

On the other hand, the gap that only gets contribution from pairs of neutral quarks has solution
\begin{equation}
 \Delta_A  \sim \frac{\Delta^{\rm CFL}_A}{2^{(7/34)}} \exp{\Big(-\frac{36}{17 x}  + \frac{21}{17}\frac{1}{x (1 + y)} + \frac{3}{2x}
\Big) } \ ,
\end{equation}
where $x \equiv g^2 \mu^2/\Lambda^2 \pi^2$, and $y \equiv
\widetilde{e}\widetilde{B}/\mu^2$ and
\begin{equation}\label{CFLgap}
\Delta_{A}^{CFL}\sim2\sqrt{\Lambda\mu-\mu^{2}}\mbox{ exp}\left(-\frac{3\Lambda^{2}\pi^{2}}{2g^{2}\mu^{2}}\right).
\end{equation}

From (\ref{mCFLgap}) it is clear that in the strong field limit, this gap increases with larger magnetic fields. If the strength of the field is such that $\tilde{e}\tilde{B}>2\mu^{2}$, we see from comparing (\ref{CFLgap}) and (\ref{mCFLgap}) that the MCFL gap surpasses the CFL gap, $\Delta_{A}^{B}>\Delta_{A}^{CFL}$.

\section{Two Quark Flavors in a Magnetic Field}

The case of two quark flavors in a magnetic field was recently considered in \cite{Topel}. Since the quarks participating in the pairing are all charged with respect to the rotated electromagnetism, the external magnetic field does not change the structure of the gap, but only its magnitude. For 2SC pairing at zero temperature, the effective thermodynamic potential with an arbitrary magnetic field is given by

\begin{eqnarray}\label{generalomega}
\Omega = \Omega_{0}+\frac{\Delta^{2}}{4G}-\frac{\mu_{db}^{4}}{12\pi^{2}}\nonumber \\\quad\quad
-\frac{eB}{4\pi^{2}}\sum_{n=0}^{[\frac{\mu_{e}^{2}}{2eB}]} \left(2-\delta_{n0}\right)\left[\frac{\mu_{e}}{2}\sqrt{\mu_{e}^{2}-2eBn}
-eBn \mbox{ ln}\left(\frac{\sqrt{\mu_{e}^{2}-2eBn}+\mu_{e}}{\sqrt{2eBn}}\right)\right] \nonumber \\
 -\frac{eB}{4\pi^{2}}\sum_{n=0}^{[\frac{\mu_{ub}^{2}}{2eB}]} \left(2-\delta_{n0}\right)\left[\frac{\mu_{ub}}{2}\sqrt{\mu_{ub}^{2}-2eBn}
 -eBn \mbox{ ln}\left(\frac{\sqrt{\mu_{ub}^{2}-2eBn}+\mu_{ub}}{\sqrt{2eBn}}\right)\right]\nonumber \quad\quad
\\
 -\frac{eB}{\pi^{2}}\sum_{n=0}^\infty \left(1-\frac{\delta_{n0}}{2}\right)\int_0^\infty e^{\frac{-\left(p_{3}^{2}+eBn\right)}{\Lambda^{2}}}\sqrt{\left(\sqrt{p_{3}^{2}+eBn}
+\overline{\mu}\right)^{2}
+\Delta^{2}}\,dp_{3}\nonumber
\\
 -\frac{eB}{2\pi^{2}}\sum_{n=0}^\infty \left(1-\frac{\delta_{n0}}{2}\right)\int_0^\infty e^{\frac{-\left(p_{3}^{2}+eBn\right)}{\Lambda^{2}}}\left|\sqrt{\left(\sqrt{p_{3}^{2}+eBn}-\overline{\mu}\right)^{2}
+\Delta^{2}}+\delta \mu \right| \,dp_{3} \nonumber
\\
 -\frac{eB}{2\pi^{2}}\sum_{n=0}^\infty \left(1-\frac{\delta_{n0}}{2}\right)\int_0^\infty e^{\frac{-\left(p_{3}^{2}+eBn\right)}{\Lambda^{2}}}\left|\sqrt{\left(\sqrt{p_{3}^{2}+eBn}-\overline{\mu}\right)^{2}
+\Delta^{2}}-\delta \mu \right| \,dp_{3}.
\end{eqnarray}
where in order to impose the neutrality conditions and to satisfy $\beta$- equilibrium constraints, we have to introduce all the chemical potentials for the conserved and commuting charges. The diagonal matrix of chemical potentials is given by

\begin{equation}
\mu_{ij,\alpha\beta}=(\mu\delta_{ij}-\mu_{e}Q_{ij})\delta_{\alpha\beta}+\frac{2}{\sqrt{3}}\mu_{8}\delta_{ij}(T_{8})_{\alpha\beta}.
\end{equation}

\vspace{8mm}

\noindent Here, $\mu$, $\mu_{e}$, and $\mu_{8}$ are the quark, electron, and color chemical potentials respectively. The generators, $Q$ and $T_{8}$, are those of the electromagnetic group $U(1)_{em}$ and the color subgroup $U(1)_{8}$. The individual quark chemical potentials then read

\begin{eqnarray}
\mu_{ur}=\mu_{ug}&=\mu-\frac{2}{3}\mu_{e}+\frac{1}{3}\mu_{8},\\
\mu_{dr}=\mu_{dg}&=\mu+\frac{1}{3}\mu_{e}+\frac{1}{3}\mu_{8},\\
\mu_{ub}&=\mu-\frac{2}{3}\mu_{e}-\frac{2}{3}\mu_{8},\label{muub}\\
\mu_{db}&=\mu+\frac{1}{3}\mu_{e}-\frac{2}{3}\mu_{8}.
\end{eqnarray}

\vspace{8mm}

\noindent By requiring 2SC quark matter to be invariant under the $SU(2)_{c}$ color subgroup, we avoid the need to introduce a second color chemical potential, $\mu_{3}$, though in general this would be present, as there are two different color charges \cite{Huang Shovkovy g2SC}. Following the notation of Ref. \cite{Huang Shovkovy g2SC}, we introduce the shorthand

\begin{eqnarray}
\overline{\mu}\equiv\frac{\mu_{ur}+\mu_{dg}}{2}=\frac{\mu_{ug}+\mu_{dr}}{2}=\mu-\frac{\mu_{e}}{6}+\frac{\mu_{8}}{3},\\
\delta\mu\equiv\frac{\mu_{dg}-\mu_{ur}}{2}=\frac{\mu_{dr}-\mu_{ug}}{2}=\frac{\mu_{e}}{2}.\label{deltamu}
\end{eqnarray}

For large magnetic fields, $eB\sim4\mu^{2}$ and assuming the density to be large enough to avoid gapless modes, the thermodynamic potential (\ref{generalomega}) reduces to

\begin{eqnarray}\label{large B omega}
\Omega=\Omega_{0}+\frac{\Delta^{2}}{4G}-\frac{\mu_{db}^{4}}{12\pi^{2}}-\frac{eB}{4\pi^{2}}\left(\frac{\mu_{e}^{2}}{2}\right)
-\frac{eB}{4\pi^{2}}\left(\frac{\mu_{ub}^{2}}{2}\right)\\
-\frac{eB}{2\pi^{2}}\int_0^\Lambda \left(\sqrt{\left(p_{3}+\bar{\mu}\right)^{2}
+\Delta^{2}}+\sqrt{\left(p_{3}-\bar{\mu}\right)^{2}
+\Delta^{2}}\right)\,dp_{3}.
\end{eqnarray}

The 2SC gap that solves the gap equation derived from this strong field potential is
\begin{equation}\label{gapequation}
\Delta=2\sqrt{\Lambda^{2}-\overline{\mu}^{2}}\exp\left( {-\frac{\pi^{2}}{2GeB}}\right) .
\end{equation}

In 2SC, all four of the participating quarks carry a charge. The total density of states is then $4[(e/2)B/4\pi^{2}]=eB/2\pi^{2}$. Defining $\bar{G}\equiv 4G$, the exponential becomes
\begin{equation}
\Delta\sim\mbox{exp}\left(-\frac{2\pi^{2}}{\bar{G}eB}\right).
\end{equation}
which is the usual BCS form.

Imposing color and electric neutralities one can show that the solutions for the chemical potentials in the strong field limit are

\begin{eqnarray}\label{mu 8 cubic}
\mu_{8}\simeq \frac{3}{2}\mu-\frac{5}{8}eB \left[\frac{9}{2\left(-12eB\mu+\sqrt{6}\sqrt{(eB)^{3}
+24(eB)^{2}\mu^{2}}\right)}\right]^{\frac{1}{3}} \nonumber
\\
+\frac{5}{8}\left[\frac{3
(-12eB\mu+\sqrt{6}\sqrt{(eB)^{3}+24(eB)^{2}\mu^{2}}}{4}\right]}^{\frac{1}{3}
\end{eqnarray}

\begin{equation}\label{mu e}
\mu_{e}=\frac{3\mu-2\mu_{8}}{5}.
\end{equation}

\section{Final Remarks}

This paper has mainly outlined the effects of a magnetic field in the gap magnitude and structure of three and two quark flavor spin zero color superconductors. Some open questions naturally emerge from these results. First, it will be interesting to investigate how the smaller number of NG bosons in the MCFL phase and the fact that none of them are charged is reflected in the comparison of the transport properties of MCFL and CFL. How will the gap parameter and chemical potentials behave at lower magnetic fields in the two-flavor case? How do the free energies of the two and three flavor systems compare when in addition to the magnetic field a small, but finite temperature is also introduced?

However, the most urgent question in the field of color superconductivity is related to the stable ground state that realizes at intermediate densities. This problem arises at intermediate densities when the Fermi surface imbalance between pairing quarks becomes too large thus giving rise to gapless modes. The imbalance itself comes from imposing beta equilibrium and neutrality or when the density is such that the strange quark mass cannot be ignored any longer. As it turns out, once the gapless modes occur, there are also chromomagnetic instabilities present, indicating one is not working in the correct ground state. A solution  to this chromomagnetic instability on which an inhomogeneous gluon condensate and an induced magnetic field are generated hence modifying the ground state of the superconductor was proposed in \cite{B-induction}. Since it involves the spontaneous generation of a rotated magnetic field by the gluon condensate, it could also serve to address some of the puzzles of the standard magnetar model \cite{SuperB}, thereby connecting magnetars and color superconductivity.

On the other hand, a magnetic field can have some other effects on the color superconductor that have not been explored yet and which could give rise to unexpected and interesting new physics. One possibility is that the Dirac structure of the gap may also include an anomalous magnetic moment term that is formed by the contraction between the spin operator and the field. If such a term breaks the same symmetry as the gap with Dirac structure $C\gamma_{5}$, it has all the right to be present once the gap is generated, since it is not protected by any symmetry after that. This reasoning is similar to the impossibility to avoid the presence of symmetric gaps in spin zero superconductivity. The need to consider a dynamical anomalous magnetic moment along with a dynamical mass was recently studied in the context of massless QED and magnetic catalysis of chiral symmetry breaking \cite{Zeeman}. Over there it gave rise to a nonperturbative Zeeman effect. It would be worthy to find out what consequences a nonperturbative anomalous magnetic moment term could have on the color superconductor and particularly, whether it can offer a different kind of solution to the intermediate density instability problem.

I am grateful to the organizers of the CSQCD conference for their invitation. This work was supported in part by the Department of Energy Nuclear Theory grant DE-FG02ER41599.

\end{document}

%% file: econfmacros.tex
\def\beq{\begin{equation}}
\def\eeq#1{\label{#1}\end{equation}}
\def\eeqn{\end{equation}}

\def\beqa{\begin{eqnarray}}
\def\eeqa#1{\label{#1}\end{eqnarray}}
\def\eeqan{\end{eqnarray}}

\let\bar=\overbar

\def\Dslash{\not{\hbox{\kern-4pt $D$}}}
\def\dslash{\not{\hbox{\kern-2pt $\del$}}}

\def\msb{{\bar{\ssstyle M \kern -1pt S}}}

\usepackage{fancyhdr,graphicx}